\documentclass[aip,rsi,reprint]{revtex4-1}

\usepackage{dcolumn}
\usepackage{bm}
\usepackage{graphicx}
\usepackage{braket}
\usepackage{booktabs}
\usepackage{array}
\usepackage{xcolor} 
\usepackage{array}
\usepackage{booktabs}
\usepackage{xcolor}
\usepackage{nicematrix}
\usepackage{multirow}
\usepackage{amssymb}
\usepackage{siunitx}

\begin{document}


\title{FIREQ: FPGA Instrumentation for
Readout and Qubit control} 




\author{Giuseppe La Capra} 
 \affiliation{Department of Electronics and Telecommunication, Politecnico di Torino, Torino, 10129, Italy.}

\author{Fabio Calabrese} 
\affiliation{Department of Electronics and Telecommunication, Politecnico di Torino, Torino, 10129, Italy.}

\author{Giorgio D'Amico} 
\affiliation{Department of Electronics and Telecommunication, Politecnico di Torino, Torino, 10129, Italy.}

\author{Christian Conti} 
\affiliation{Department of Electronics and Telecommunication, Politecnico di Torino, Torino, 10129, Italy.}
\author{Andrea De Simone} 

\affiliation{Department of Electronics and Telecommunication, Politecnico di Torino, Torino, 10129, Italy.}
\author{Deborah~Volpe} 
\affiliation{Istituto Nazionale di Geofisica e Vulcanologia, Roma, 00143, Italy.}

\author{Angelo~Nucciotti} 
\affiliation{Department of Physics, University of Milano-Bicocca, Milano, 20126, Italy.}
\affiliation{INFN - Milano Bicocca, Milano, 20126, Italy}
\affiliation{Bicocca Quantum Technologies (BiQuTe) Centre, Milano, 20126, Italy}

\author{Rodolfo~Carobene} 
\affiliation{Department of Physics, University of Milano-Bicocca, Milano, 20126, Italy.}
\affiliation{INFN - Milano Bicocca, Milano, 20126, Italy}
\affiliation{Bicocca Quantum Technologies (BiQuTe) Centre, Milano, 20126, Italy}

\author{Claudio~Gatti} 
\affiliation{INFN - Laboratori Nazionali di Frascati - 00044 Frascati (RM), Italy}

\author{Andrea~ Giachero} 
\affiliation{Department of Physics, University of Milano-Bicocca, Milano, 20126, Italy.}
\affiliation{INFN - Milano Bicocca, Milano, 20126, Italy}
\affiliation{Bicocca Quantum Technologies (BiQuTe) Centre, Milano, 20126, Italy}

\author{Fabrizio~Riente} 
\affiliation{Department of Electronics and Telecommunication, Politecnico di Torino, Torino, 10129, Italy.}
\affiliation{INFN - Torino, Torino, 10125}


\date{\today}

\begin{abstract}
We present FIREQ (FPGA Instrumentation for Readout and Qubit control), an open-source RFSoC-based framework for the control and readout of superconducting qubits. FIREQ combines a modular AXI-compliant firmware architecture with a PYNQ-based software stack designed to support extensible hardware integration, deterministic experiment timing, and low-overhead execution of repeated calibration and characterization workflows. The firmware implements direct RF synthesis and acquisition, trigger-based sequencing, programmable pulse generation, frequency-multiplexed readout, and memory-efficient acquisition and waveform buffering. The software adopts a client–server architecture with streamed data transfer and dependency-aware configuration updates to reduce host–device and reconfiguration overhead during parameter sweeps. On an AMD Zynq UltraScale+ RFSoC ZCU216, FIREQ generates RF pulses up to \SI{9.3}{\giga \hertz} with a pulse-duration resolution of  \SI{107}{\pico \s} and an event-timing resolution of \SI{1.7}{\nano \s}. FPGA resource utilization is compared with representative open-source RFSoC control frameworks, showing a low BRAM footprint while retaining full-rate I/Q generation and acquisition. The RF output is characterized in terms of phase noise, noise spectral density, and inter-channel timing skew. End-to-end operation is validated on a superconducting qubit through resonator spectroscopy, Rabi, Ramsey, and relaxation measurements, yielding $T_1 =$ \SI{6.94}{\micro \s} and $T_2^* =$ \SI{13.50}{\micro \s}. FIREQ can therefore be used both as a qubit-control platform and as an experimental environment for evaluating alternative control and readout IP architectures.
\end{abstract}


\maketitle 


\section{Introduction}
As superconducting quantum processors scale to larger qubit counts, the classical infrastructure required to control and measure quantum states becomes an increasingly relevant constraint alongside qubit coherence and fidelity \cite{brennan2025classical, ahmad2022scalable, scalability}. In superconducting platforms, this infrastructure operates at the interface between cryogenic quantum hardware and room-temperature electronics and must handle microwave signals under stringent timing, spectral, and synchronization requirements.
\begin{figure*}[ht]
	\centering
	\includegraphics[width=1.00\linewidth]{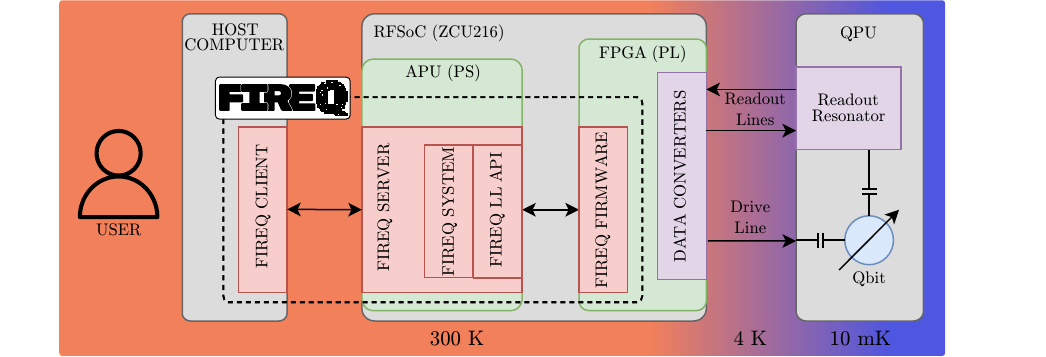} 
\caption{System overview: a room-temperature host drives an RFSoC board, composed of a Processing System (PS), Programmable Logic (PL) and Data Converters that generates and acquires the qubit drive and readout signals across the \SI{300}{\kelvin}-\SI{10}{\milli \kelvin} stages. The FIREQ framework, whose components are colored in red, comprises the host and device software (client and server) and the device firmware.}
\label{fig:GA}
\end{figure*}
Superconducting qubits are controlled and read out through microwave pulses in the gigahertz range, requiring the generation, acquisition, and processing of radio frequency (RF) signals with sub-nanosecond timing accuracy and high spectral purity. As qubit counts increase, the required number of control and readout channels, aggregate data throughput, and synchronization complexity increase accordingly. 
In addition, the limited signal-to-noise ratio of dispersive readout, combined with the need for repeated shot acquisition, further imposes high-throughput data acquisition and efficient signal processing pipelines. 
The RF hardware is only part of the scaling problem. Experiment throughput also depends on how efficiently a high-level experiment description is translated into hardware operations. 
Calibration, characterization, and algorithm execution involve repeated pulse sequences, multidimensional parameter sweeps, and large volumes of acquired I/Q data, making software–hardware interaction and data movement part of the effective experiment latency.  
The effective performance of a control platform therefore depends on the underlying RF/FPGA hardware, the specifications of its Analog-to-Digital Converters (ADCs) and Digital-to-Analog Converters (DACs), the efficiency of the software-hardware interface and data-transfer pipeline, and integration with higher-level software.

Conventional superconducting-qubit control setups typically combine arbitrary waveform generators, analog mixers, digitizers, and external synchronization hardware. Although such instrumentation provides high analog performance and flexibility, scaling to larger channel counts increases hardware complexity, calibration effort, cabling, and cost.
FPGAs and Radio Frequency System-on-Chip (RFSoC) platforms provide an alternative in which waveform generation, acquisition, digital signal processing, and timing control can be integrated within a programmable architecture. 
Frameworks such as the Quantum Instrumentation Control Kit (QICK) \cite{stefanazzi2022qick} have demonstrated pulse-level control and readout using direct RF synthesis and acquisition on RFSoC devices.

Existing open RFSoC frameworks provide flexible pulse-level control and readout, but high-repetition workflows may still incur non-negligible overhead from host–device interaction, repeated configuration, and data movement. In addition, extensibility at both the firmware and software levels becomes relevant when custom signal-processing IPs or higher-level quantum software interfaces must be integrated.

Here we present FIREQ (FPGA Instrumentation for Readout and Qubit control), an RFSoC-based control and readout framework for superconducting qubits. FIREQ adopts a hardware-software co-design approach, integrating signal generation, acquisition, and processing within a single programmable platform (Fig.~\ref{fig:GA}). 
The firmware is organized as a set of modular AXI-compliant peripherals for timing, signal generation and acquisition. A PYNQ-based \cite{AMD_PYNQ_2025} software stack provides peripheral abstraction, hierarchical configuration, and dependency-aware updates, allowing new IP cores to be integrated without restructuring the upper software layers. FIREQ is released as an open-source framework, including the client software,
the server-side software and executable RFSoC bitstream, and the user
documentation, available through the project documentation site \url{https://fireq-docs.polito.it} and archived
through versioned Zenodo records~\cite{fireq_client_zenodo,
fireq_server_zenodo, fireq_docs_zenodo}.




A trigger-based timing engine schedules hardware operations deterministically, while dedicated DSP pipelines implement modulation, demodulation, decimation, and acquisition processing on the programmable logic. 
By exploiting the embedded ADCs and DACs of RFSoC platforms for direct digital synthesis and acquisition, FIREQ removes the need for external I/Q modulation and demodulation stages and reduces system latency. 
Experiment execution is coordinated through a client–server architecture that separates high-level experiment definition from device-side configuration and acquisition. The server supports concurrent command handling and streamed acquisition, while dependency-aware configuration updates reduce recomputation during parameter sweeps.

We characterize the RF output in terms of phase noise, noise spectral density, and inter-channel timing skew, compare FPGA resource utilization with representative open-source RFSoC frameworks, and validate end-to-end operation through standard calibration and coherence measurements on a superconducting qubit. 

The primary contributions of this work are summarized as follows:
\begin{itemize}
    \item \textbf{FIREQ framework:} We introduce a flexible and expandable qubit control and readout framework based on AXI-compliant peripherals, designed to reduce hardware resource utilization and simplify system reconfiguration.
    \item \textbf{Envelope interpolation technique:} We introduce an on-the-fly interpolation method that reconstructs pulse envelopes from a reduced set of reference samples, decreasing waveform-memory requirements and enabling envelope reuse.
    \item \textbf{Memory-efficient acquisition buffering:} We map decimated and accumulated acquisition buffers to UltraRAM, reserving higher-bandwidth BRAM resources for datapaths with stricter throughput requirements.
    \item \textbf{Integrated software stack:} We develop an accompanying modular software stack that supports HW reconfiguration with minimal changes and eases the execution of experiments through standardized hierarchical configuration and dependency resolution systems.
    \item \textbf{Comprehensive RF performance characterization and qubit-level validation:} We characterize phase noise, noise spectral density, and inter-channel timing skew and validate end-to-end operation through standard calibration and coherence measurements on a superconducting qubit.
\end{itemize}

\section{Background}
\subsection{Qubit states}
\begin{figure}[htbp]
    \centering
    \includegraphics[width=0.25\textwidth]{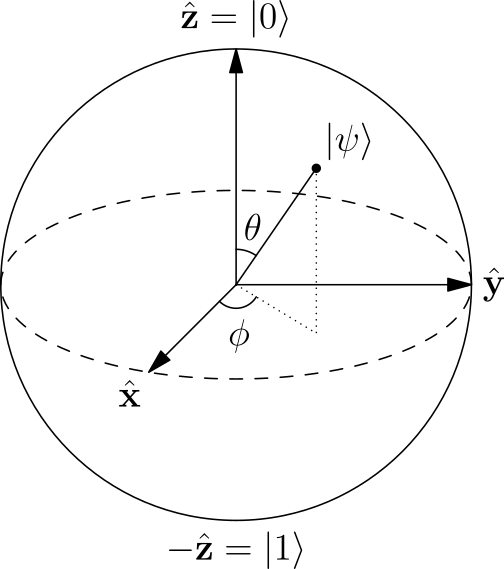}
    \caption{Bloch Sphere Representation. Image by Glosser.ca (Wikimedia Commons), used under CC BY-SA 3.0.}
    \label{fig:bloch_sphere}
\end{figure}

Qubits are quantum systems characterized by discrete energy levels. By virtue of quantum superposition, a qubit can occupy a coherent superposition of its energy eigenstates. The state can therefore be represented as a vector expressed as a complex linear combination of a set of $N$ orthonormal basis states:
\begin{equation}
    |\psi\rangle = \sum_{i=0}^{N-1} C_i \ket{i} = C_0\ket{0} + C_1\ket{1} + \dots + C_{N-1}\ket{N-1}
\end{equation}
where $C_x \in \mathbb{C}$.
The squared magnitude of each complex coefficient ($P_x=|C_x|^2$) gives the probability of measuring the corresponding basis state.
Although superconducting circuits exhibit multiple energy levels, conventional qubit operation is restricted to the two lowest levels. In conventional qubit operation, only the ground and first excited states, denoted by $\ket0$ and $\ket1$, are used. These states form the computational basis, consequently, the state vector $\vert{}\psi\rangle$ is defined as:
\begin{equation}
    |\psi\rangle = \alpha |0\rangle + \beta |1\rangle, \quad \text{where } \alpha, \beta \in \mathbb{C} \text{ and } |\alpha|^2 + |\beta|^2 = 1
    \label{eq:qubit_state}
\end{equation}
Qubit control, commonly referred to as qubit driving, modifies the amplitudes and phases defining the quantum state. In multi-qubit systems, coherent control combined with entangling interactions enables the implementation of quantum algorithms.
Qubit measurement, or readout, is equally important and projects the quantum state onto one of the measurement basis states. Experiments are therefore repeated over multiple shots to estimate the probabilities associated with the $\ket0$ and $\ket1$ outcomes. 
Additionally, the state vector is usually represented graphically on a unit sphere called the Bloch sphere, where the two basis states lie on the z axis and have opposite sign, as shown in Fig.~\ref{fig:bloch_sphere}.
In the Bloch sphere vector space, Equation~\ref{eq:qubit_state} for the qubit state becomes:
\begin{equation}
\ket{\psi} = \cos\left(\frac{\theta}{2}\right)\ket{0} + e^{i\phi}\sin\left(\frac{\theta}{2}\right)\ket{1}
\label{eq:bloch_state}
\end{equation}
where the polar angle $\theta$ is the variable that determines the readout probabilities of the two states, whereas the azimuth $\phi$ is the relative phase between the two coefficients. Passing from Equation~\ref{eq:qubit_state} to Equation~\ref{eq:bloch_state} factors out a global phase, but because an overall phase shift is physically unobservable and yields no effect on quantum operations or measurement statistics, it can be safely discarded. Likewise, a phase shift applied at the start of an experiment has no measurable outcome.

\subsection{Control and readout of superconducting qubits}
Superconducting qubits are coherently controlled by resonant microwave pulses, which induce rotations of the state vector about axes in the equatorial plane of the Bloch sphere. Rotation about orthogonal axes in the equatorial plane can be implemented by changing the phase of the RF pulses by 90°.
Quadrature (IQ) modulation is commonly employed to generate these pulses, either through external analog mixing or by direct digital synthesis followed by Radio Frequency Digital to Analog Conversion (RF-DAC).
The resulting rotation angle depends on the pulse amplitude and duration. Pulse envelopes are therefore calibrated to implement the desired rotation. A calibrated control pulse implements a quantum gate. For example, an $X_{\pi}$ pulse implements a 180° rotation about the x-axis, whereas an $X_{\pi/2}$ pulse implements a 90° rotation.
Maintaining a well-defined phase relationship between the IQ modulation tones and the qubit is essential for coherent state control. Phase offsets correspond to rotations of the control reference frame about the z-axis. This principle can be exploited to implement z-axis rotations without generating an additional physical pulse. These operations are commonly referred to as virtual-Z gates\cite{Z_gates}.
Dispersive readout is performed by injecting a modulated probe pulse into a two-port resonator capacitively coupled to the qubit and acquiring the transmitted signal with an ADC. In the dispersive regime, the resonator response depends on the qubit state. As a result, the measured scattering parameter $S_{21}$ becomes state dependent. The qubit state can therefore be inferred from the amplitude and phase response of the transmitted readout signal. 
Superconducting-qubit control and readout therefore require the generation and acquisition of IQ-modulated RF signals with precise control of amplitude, phase, waveform shape, and timing.
Meeting these requirements with conventional laboratory instrumentation can be challenging. Architectures based on arbitrary waveform generators (AWGs), external mixers, and trigger distribution networks are bulky, expensive and may suffer from I/Q imbalance and synchronization overhead. For these reasons, integrated FPGA/RFSoC-based platforms have emerged as an attractive solution, offering compact form factors, deterministic digital processing pipelines and improved flexibility.
%
\subsection{Related work}
\begin{table*}[]
  \centering
  \caption{Comparison of representative FPGA/RFSoC-based control
  and readout frameworks for superconducting qubits.
  The table reports the most recent implementations considered in these works.}
  \label{tab:related}
  \renewcommand{\arraystretch}{1.3}
  \footnotesize
  \begin{NiceTabular}{
    @{}c c c c >{\centering\arraybackslash}m{5.0cm}@{}
  }[
    code-before = \rowcolor{yellow!20}{9}
  ]
    \hline
    \hline
    \textbf{Framework}
    & \textbf{Platform}
    & \textbf{Abstraction}
    & \textbf{Multi-board}
    & \textbf{Key focus} \\
    \hline
    \hline

    QICK \cite{stefanazzi2022qick}
    & RFSoC (ZCU111/216)
    & Pulse / Gate
    & Yes
    & Reference open-source architecture, implements a time-critical
      soft-core to handle qubit manipulation. \\
    \hline

    ARTIQ \& Sinara \cite{artiq}
    & PXIe expansion boards
    & Real-time kernels
    & Yes
    & Co-designed open-source framework for quantum information
      experiments. \\
    \hline

    QubiC \cite{qubic,qubic-distributed}
    & RFSoC (ZCU216)
    & Pulse / Gate
    & Yes
    & Open-source stack with custom soft-core processor and dedicated
      Instruction Set Architecture (ISA). \\
    \hline

    RISC-Q \cite{risc-q}
    & RFSoC (ZCU216)
    & Pulse
    & --
    & Qubit control system compliant with the RISC-V environment,
      with custom ISA expansion for qubit control. \\
    \hline

    HiSEP-Q 2.0 \cite{guo2023hisep}
    & RFSoC (ZCU216)
    & Pulse
    & Yes
    & Based on the Vicuna \cite{vicuna} vector coprocessor for RISC-V
      cores to control up to 128 qubits with a single custom instruction. \\
    \hline

    YAQCS-arch \cite{yaqcs-arch}
    & PXIe expansion boards
    & Pulse / Gate
    & --
    & RISC-V architecture for a compact qubit control system exploiting
      standard RISC-V instructions. \\
    \hline

    Qibo \cite{qibo_paper}
    & Hardware-agnostic
    & Gate
    & --
    & High-level abstraction stack supporting multiple backends. \\
    \midrule
    \hline

    \textbf{FIREQ (this work)}
    & RFSoC (ZCU216)
    & Pulse
    & Planned
    & Peripheral-based qubit control system with a standard AXI interface
      and a complete remote software stack. \\
    \hline
    \hline
  \end{NiceTabular}
\end{table*}
Over the last decade, advances in superconducting qubits fabrication have driven researchers to develop advanced control systems based on commercially available FPGAs. This section outlines representative qubit control platforms based on FPGA/RFSoC, and summarizes their main characteristics in Table~\ref{tab:related}.
These solutions provide the flexibility and computational capabilities required to generate, acquire, and process the signals involved in qubit control and readout. 
Indeed, an initial step toward the control of qubits was proposed by the Advanced Real-Time Infrastructure for Quantum Physics (ARTIQ) and Sinara\cite{artiq} through a general open-source software and hardware stack for quantum physics experiments. Sinara provides modular hardware that can be configured according to the target application and is controlled through the ARTIQ software framework. The combination of Sinara and ARTIQ provides nanosecond-level timing resolution and sub-microsecond latency, as time-critical kernels are executed on the FPGA. ARTIQ further exposes FPGA functionality through a high-level Python interface.\\
QubiC \cite{qubic} is another FPGA-based radio frequency control platform for qubit manipulation. 
Its second major release supports both gate- and pulse-level workflows. The gate-level abstraction allows users to define experiments without directly specifying pulse-level waveforms. The current implementation targets the AMD Zynq UltraScale+ ZCU216 RFSoC platform and includes a custom analog front-end.
The QubiC software stack is distributed between the embedded processor of the SoC through Python interface and an external host. More recently, an extension for synchronizing multiple FPGAs has been released \cite{qubic-distributed}, enabling distributed qubit control and scaling the system across multiple boards. \\
The Quantum Instrumentation Control Kit (QICK) \cite{stefanazzi2022qick} is a reference open-source FPGA-based architecture for the control and readout of superconducting qubits. It is designed to provide a low-latency solution for control and readout of superconducting qubits.
It relies on the implementation of a soft-core processor, the T-Processor, which handles the time-critical kernels required for qubit control. It integrates several peripherals for waveform generation and digital signal processing and communicates with the host through the Processing System (PS), the processor subsystem embedded in the system-on-chip (SoC).
By exploiting the capabilities of the RFSoC family, QICK enables direct digital synthesis and acquisition in the microwave domain without requiring external instrumentation.
A key feature of QICK is its ability to implement pulse sequences and measurement protocols with deterministic timing and reduced latency, which is essential for advanced quantum control tasks such as active qubit reset, feedback-based error mitigation, and adaptive experiments. Additionally, through the PS, QICK provides a Python-based interface that simplifies experiment construction while maintaining fine-grained access to hardware-level operations. 
Subsequent developments have extended QICK toward multi-board and distributed operation.
It currently supports multiple AMD Zynq UltraScale+ platforms (RFSoC ZCU111, RFSoC ZCU216, and RFSoC4x2), which can also be interconnected to perform distributed and synchronized qubit control, as proposed in XCOM \cite{xcom2026} or in Manarat\cite{manarat}.
Several works have explored RISC-V extensions as a route toward standardized instruction-set architectures for qubit control.

In particular, RISC-Q \cite{risc-q} introduces the concept of a Quantum Control System-on-Chip (QCSoC), moving from the versatility of FPGA solutions toward the efficiency of custom-designed hardware. The proposed architecture can control different qubit technologies, including superconducting, neutral-atom, and trapped-ion qubits, through a collection of RF generators and RF decoders interfacing with the cryostat. 
Furthermore, the integration of specialized accelerators with the RISC-V core enables fast signal pre- and post-processing in the classical domain.
Along similar lines, Yet Another Quantum Computing Suite (YAQCS-arch)\cite{yaqcs-arch} adopts a modular architecture for a qubit controller based on the flexibility provided by the PCI Express eXtensions for Instrumentation (PXIe) and the standard RISC-V ISA, connecting the RF front-end to the main RISC-V controller with an ISA extension to support qubit control both at pulse and gate manipulation level, defining a set of quantum instructions that can be expanded into the well defined RISC-V ISA. 
Another example is HiSEP-Q 2.0 \cite{guo2023hisep}, a RISC-V vector-extension architecture for scalable qubit systems. It builds upon its previous implementation by moving control operations to a RISC-V Vector (RVV) engine. By exploiting the characteristics of the RVV engine and introducing custom extensions for qubit control, the architecture enables the concurrent control of up to 128 qubits with a single instruction.
These architectures reflect a shift from deployment-oriented laboratory controllers such as QICK toward ISA- and microarchitecture-oriented quantum-control processors.
\\
In parallel, other research efforts have focused on raising the abstraction level from pulse-level control, which requires explicit waveform specification, to gate-level interfaces closer to the representation of quantum circuits.
An example of this approach is Qibo \cite{qibo_paper}, a software stack capable of interfacing with different control platforms, such as QICK. Qibo abstracts away hardware-specific details and the waveform definitions required for qubit control, thereby facilitating rapid prototyping of quantum circuits.


\section{Motivations and Design Rationale} \label{sec:design_rationale}
The scaling of superconducting-qubit systems is strongly constrained by the number of physical interconnects between the quantum processor and its control electronics\cite{scaling_interconnects}. This motivates more compact control architectures and, in the longer term, the integration of control electronics closer to the cryogenic quantum hardware\cite{cryogenic_control}.
Such integration, however, introduces stringent constraints on power consumption and heat dissipation, particularly at low-temperature stages where the available cooling power is limited \cite{kawabata2026}.
Accordingly, scalability is determined not only by the number of RF input/output channels that a platform can support, but also by the hardware cost required to implement each control and readout channel. Resource-efficient architectures are therefore of interest both for current room-temperature instrumentation and for the development of future integrated control systems.
FIREQ is designed with this objective in mind. In addition to providing a complete room-temperature FPGA-based control and readout platform, FIREQ serves as a testbed for the development and evaluation of resource-efficient mixed-signal processing IPs. Its modular peripheral-based architecture enables individual hardware blocks to be integrated, replaced, and evaluated with limited impact on the rest of the system.
Both the hardware and software stacks were co-designed for extensibility. On the hardware side, custom peripherals expose standard AXI4 interfaces. On the software side, Python drivers and a system-level abstraction layer provide a uniform mechanism for integrating new IP blocks.
The overall design prioritizes low FPGA resource utilization while maintaining signal quality, low latency, and fine timing resolution.

\section{FIREQ System Architecture}
\begin{figure*}[ht!]
	\centering
	\includegraphics[width=1\linewidth]{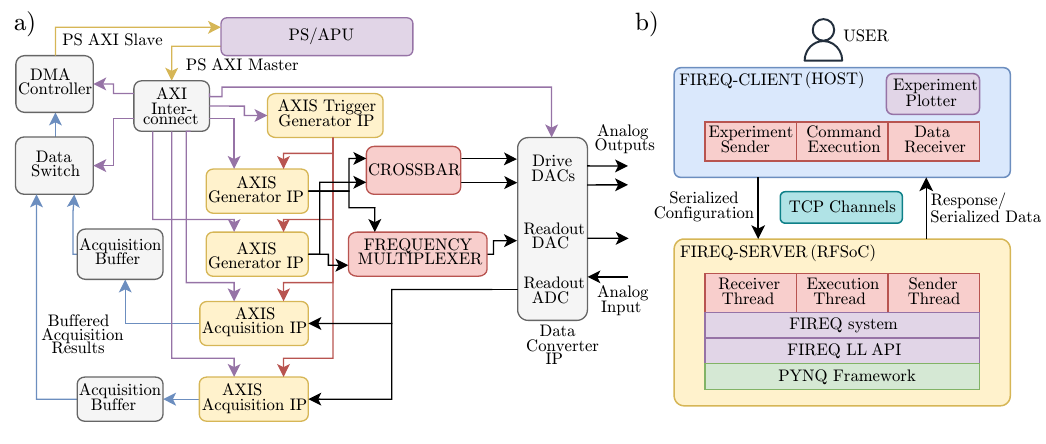} 
\caption{a) overview of the FIREQ Firmware, showing the custom peripherals in yellow that handle timing, signal generation and acquisition, and how they are connected to the RFSoC's APU (PS) through AXI connections; b) the client-server software architecture, showing the inbound and outbound configuration and data channels and the PYNQ-based software abstraction stack.}
\label{fig:FIREQ_firmware_and_client_server}
\end{figure*}
The FIREQ system consists of three main components:
\begin{enumerate}
    \item Firmware/bitstream/overlay: the RTL-based hardware design implemented in the FPGA. The bitstream refers to the binary file generated by Vivado, whereas the overlay comprises the bitstream and the associated hardware hand-off files, used by PYNQ to dynamically configure the FPGA.
    \item Server: a Python application running on the RFSoC that translates client requests into firmware configuration and execution commands and returns acquired data over the network
    \item Client: a Python application running on the host machine that provides the user-facing interface for experiment definition and execution.
\end{enumerate}

\subsection{Firmware}
The firmware is organized as a set of peripherals connected to the Application Processing Unit (APU), which performs experiment configuration prior to execution, as shown in Fig.~\ref{fig:FIREQ_firmware_and_client_server}(a). Each experiment is fully specified before execution through the use of memories and registers associated with the individual peripherals. All peripherals expose AXI4-compatible interfaces and are memory-mapped into the APU address space.
Custom peripherals handle signal generation, acquisition and timing colored in yellow, pulse routing and frequency multiplexing colored in red. All other peripherals are provided by AMD and they either handle the AXI infrastructure or are hard IPs, such as the PS and the data converter IP, the latter of which exposes the digital data connections of the RF DACs and ADCs.
Currently, two signal generation and acquisition peripherals are instantiated in the design, powering the control of up to two qubits. RF DACs and ADCs work close to their maximum sampling frequency (9.34 and 2.33 GSps respectively), allowing for a high instantaneous signal bandwidth and fine time resolution with no interpolation or decimation used to reduce the working throughput of the signal generation and acquisition IPs.

\subsubsection{Timing engine}
Experiment timing is controlled by a custom trigger generator IP. This peripheral generates up to 15 parallel \textit{triggers}, each asserted for a single clock cycle to schedule time-critical events. Triggers are divided into two categories: drive and readout. Drive triggers are routed exclusively to the signal generators, whereas readout triggers are distributed to both generation and acquisition IPs to synchronize readout-pulse generation and acquisition. 
The trigger sequence is stored in an internal memory containing up to 8096 entries. Each entry specifies a relative time delay and the set of triggers to be asserted when the delay expires.
Each timing event can generate up to 15 parallel drive or readout triggers, with a maximum delay of approximately 7.35 seconds. If a longer delay is needed, multiple ones can be concatenated.
The timing engine also generates shot- and experiment-level start/stop signals, allowing connected peripherals to track experiment execution boundaries.

\subsubsection{Signal Generation}
\begin{figure}[ht]
    \centering
    \includegraphics[width=1.0\linewidth]{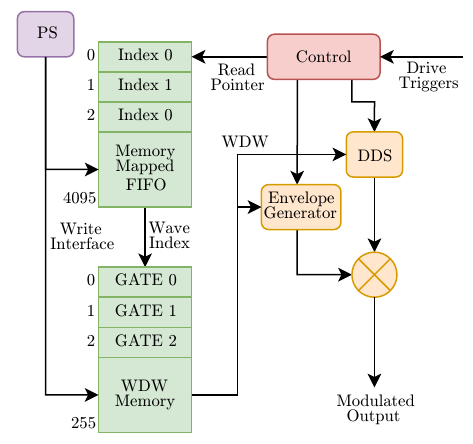}
    \caption{Simplified block diagram of the sequencing mechanism of drive pulses present in the FIREQ signal generator IP. The memory mapped FIFO indexes the WDW memory, which in turn stores control signals to generate a pulse. In this example, the drive wave order is: GATE 0, GATE 1, GATE 0.}
    \label{fig:gen_sequencer}
\end{figure}
RF pulse generation is implemented by a custom peripheral combining a programmable envelope lookup table with a Direct Digital Synthesis (DDS) engine for carrier generation (sine and cosine). 
Each generator can produce a sequence of phase-coherent drive pulses at a programmable carrier frequency, followed by a readout pulse with independently configurable frequency and phase. Drive and readout pulses share the same generator output and are subsequently routed to the appropriate DACs or downstream IP blocks according to pulse-level control data. Unlike architectures that employ separate generator blocks for drive and readout, for example QubiC uses independent generators, whereas QICK includes dedicated blocks for multiplexed readout, FIREQ reuses the same generator together with downstream multiplexing IPs. This organization increases routing flexibility while reducing the number of generator instances required per qubit.
Moreover, this scheme simplifies the management of the drive and readout modulation tones, which are programmed through specific registers before the start of the experiment and do not require per-pulse reconfiguration or runtime computation to preserve phase coherence. This differs from the approaches in which phase-related parameters are updated dynamically during pulse execution like QICK. 
Each pulse is encoded by a 128-bit Wave-Definition Word (WDW). Each generator stores up to 256 such words in an internal WDW memory. Each 128-bit word specifies the control fields required either to generate a pulse or to update the carrier phase for a virtual-Z operation.
As shown in Fig.~\ref{fig:gen_sequencer}, this memory is then indexed by a memory-mapped FIFO providing the order with which pulses are executed. The memory-mapped FIFO contains up to 4096 indices and the APU accesses this structure as a memory-mapped buffer, whereas the generator consumes one entry at each drive trigger, starting from address zero. 
The read pointer is reset at the start of an experiment shot.
This design encapsulates pulse sequencing entirely within the generator IP and avoids the need for a soft-core processor in the time-critical generation path.
The readout pulse is also defined by a wave definition word but it is programmed to a register, and can also be used to generate pulses manually, either on the drive or readout outputs.

\subsubsection{Signal crossbar and frequency multiplexing}
The signal generator output includes pulse-level routing metadata that specifies the destination of each pulse. 
Each drive pulse can be routed through the crossbar to a programmable subset of drive DACs. 
The subset is defined within the wave definition word. 
Readout waveforms are directly routed to a frequency-multiplexing IP that sums the output of multiple generators onto a common readout channel. Pulse type (readout/drive) and routing (DAC subset) information are encoded in a specific set of signals defined within the AXI Stream interface protocol (TUSER). The current implementation also propagates the routing field for readout pulses, although this field is not yet used. It is reserved for future support of multiple readout output lines.
To reduce the number of resources used for frequency multiplexing, DSP resources are configured in SIMD mode to fully utilize their native parallelism.

\subsubsection{Signal Acquisition}
The acquisition IPs perform signal demodulation and framing and support both accumulation and decimation modes for the data output. Dedicated registers configure the demodulation frequency, initial phase, acquisition delay, and acquisition-window duration. 
Furthermore, to compensate for the time-of-flight of the signal and the latency of the generation chain, the acquisition IP can also wait for a number of cycles after the trigger has arrived before starting to acquire the signal.
The acquisition IP has two outputs, one is a full-rate raw I/Q stream, providing the demodulated signal, and the other is a decimated/accumulated output, which can be programmed to provide either the decimated version of the demodulated signal or the accumulated raw samples. Decimated samples provide a low-pass filtered version of the demodulated signal with a lower memory footprint, while the accumulated values provide a measurement of the energy/power of the acquired pulse.

\subsubsection{Memory-efficient acquisition buffering}
On-chip Block RAM (BRAM) capacity constrains the number of memory-intensive peripherals that can be instantiated in an FPGA design.
Both I/Q generation and acquisition require high-bandwidth on-chip memories for envelope storage, DDS lookup tables, and acquired waveform buffers.
Reducing the utilization of such resources is critical to improve the resource-efficiency of the system.
UltraScale+ FPGAs also provide UltraRAM (URAM) resources, which offer larger storage capacity per block than BRAMs, albeit with different timing characteristics. 
In FIREQ, the acquisition IPs leverage these resources to store the accumulated/decimated values, leaving BRAM resources where higher throughput is necessary. One URAM block is used for each acquisition IP to store up to 8192 decimated I/Q samples (equivalent to a 14 $\mu$ acquisition window) or 4096 accumulated I/Q values, equivalent to 8 BRAM blocks.
Raw acquisition data, due to the higher speed necessary, is instead stored into BRAM buffers able to contain 2048 I/Q samples, using 2 BRAMs per buffer.
Among the implementations considered in Table~\ref{tab:area_comparison}, the compared designs rely primarily on BRAM resources for comparable buffering functions, even when providing accumulated or decimated samples to be extracted by the APU (like in QICK).

\subsubsection{Memory-efficient signal generation}
Conventional LUT-based pulse generators require substantial on-chip memory to store both pulse envelopes and carrier lookup tables for DDS-based modulation.
Different approaches can be used to reduce the number of resources used, especially BRAMs, such as using CORDIC-based sine and cosine generation for the modulating tones and using AM modulation instead of full I/Q modulation (like in RISC-Q), or relying on the interpolating filters present in the RFSoC DACs to reduce the digital throughput (like in some generators in QICK). However, these solutions reduce the flexibility and quality of the signals generated, and, in the case of the CORDIC approach, greatly increase latency.
\begin{figure}[ht]
    \centering
    \includegraphics[width=0.95\linewidth]{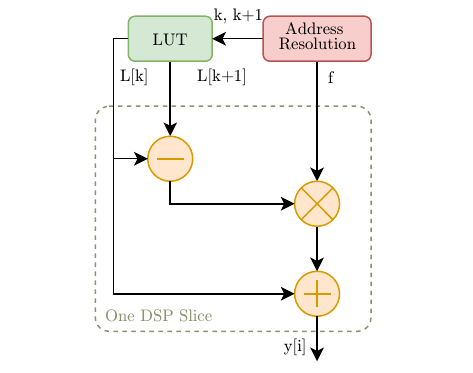}
    \caption{Block diagram of the envelope interpolator, showing the mathematical operations that are mapped to a single DSP Slice.}
    \label{fig:envelope_interpolator}
\end{figure}
\begin{table*}[!t]
  \centering
  \caption{Post-implementation resource utilization of
  FPGA/RFSoC-based control frameworks for superconducting qubits.
  All reported implementations target the AMD Zynq UltraScale+
  RFSoC ZCU216.}
  \label{tab:area_comparison}
  \renewcommand{\arraystretch}{1.3}
  \setlength{\tabcolsep}{4.5pt}
  \footnotesize

  \begin{NiceTabular}{
    @{}l
    r
    r
    r
    r
    r
    r
    r
    r@{}
  }
    \hline
    \hline

    \textbf{Framework}
    & \textbf{Total LUTs}
    & \textbf{Logic LUTs}
    & \textbf{LUTRAMs}
    & \textbf{SRLs}
    & \textbf{FFs}
    & \textbf{BRAM}
    & \textbf{URAM}
    & \textbf{DSPs} \\
    \hline
    \hline

    QICK \cite{stefanazzi2022qick}
    & 141\,348
    & 113\,050
    & 7\,650
    & 20\,648
    & 243\,829
    & 877.5  
    & 0
    & 2\,049 \\
    \hline

    QubiC \cite{qubic,qubic-distributed}
    & 51\,665
    & 38\,463
    & 1\,068
    & 12\,134
    & 102\,885
    & 350
    & 0
    & 1\,196 \\
    \hline

    RISC-Q \cite{risc-q}
    & 71\,930
    & 70\,131
    & 1\,168
    & 631
    & 147\,576
    & 272 
    & 0
    & 984 \\
    \hline

    \rowcolor{yellow!20}
    \textbf{FIREQ (this work)}
    & 32\,774
    & 27\,796
    & 552
    & 4\,426
    & 57\,992
    & 132.5 
    & 2
    & 380 \\
    \hline
    \hline
  \end{NiceTabular}

  \vspace{1mm}
  \begin{minipage}{0.99\textwidth}
    \footnotesize
    \textit{Note:} QICK, QubiC, RISC-Q and FIREQ resource utilization refers to their
    complete post-route RFSoC designs. We were unable to build the firmware for
    HiSEP-Q 2.0 from its public repository.
  \end{minipage}
\end{table*}
\begin{table*}[!t]
  \centering
  \caption{Resource usage normalized per qubit and bandwidth ($\textit{metric}*\text{qubit}^{-1}*\text{GHz}^{-1}$). Only includes drive/readout generation and readout acquisition subsystems, ignoring the control overhead and acquisition buffers.}
  \label{tab:per_qubit_area}
  \renewcommand{\arraystretch}{1.3}
  \setlength{\tabcolsep}{4.5pt}
  \footnotesize

  \begin{NiceTabular}{
    @{}l
    r
    r
    r
    r
    r
    r
    r
    r@{}
  }
    \hline
    \hline

    \textbf{Framework}
    & \textbf{Total LUTs}
    & \textbf{Logic LUTs}
    & \textbf{LUTRAMs}
    & \textbf{SRLs}
    & \textbf{FFs}
    & \textbf{BRAM}
    & \textbf{URAM}
    & \textbf{DSPs} \\
    \hline
    \hline

    QICK \cite{stefanazzi2022qick}
    & 1679.1
    & 1410.1
    & 32.7
    & 236.3
    & 3863.8
    & 15.5
    & 0
    & 32.7 \\
    \hline

    RISC-Q \cite{risc-q}
    & 2145.5
    & 2126.5
    & 0
    & 19.0
    & 4471.5
    & 8
    & 0
    & 37 \\
    \hline

    \rowcolor{yellow!20}
    \textbf{FIREQ (this work)}
    & 1631.9
    & 998.9
    & 21.2
    & 611.8
    & 4233.6
    & 8.8
    & 0
    & 39.0 \\
    \hline
    \hline
  \end{NiceTabular}
  \vspace{1mm}
  \begin{minipage}{0.99\textwidth}
    \footnotesize
    \textit{Note:} While total FPGA resource utilization is reported for QubiC, per-qubit resources could not be estimated due to its use of a custom Python-based meta-HDL framework that programmatically generates HDL code, obscuring per-channel hierarchy boundaries and making static per-qubit logic isolation non-trivial.
  \end{minipage}
\end{table*}
To reduce envelope-memory requirements, FIREQ allows multiple pulse operations to reuse the same stored envelope to achieve different X and Y rotations.
Additional pulse-level controls include programmable digital attenuation and I/Q swapping, enabling different rotation amplitudes and axes to be generated from the same stored envelope.
FIREQ further implements an on-the-fly linear interpolation mechanism that reconstructs pulse-envelope samples from a reduced set of stored reference samples. This decouples the number of stored samples from the temporal duration of the generated pulse, reducing envelope-memory requirements.
This system, shown in Fig.~\ref{fig:envelope_interpolator}, is based on DDS with first-order Taylor correction and it works by first calculating the address using a start offset and an increment, followed by the LUT access of two I/Q samples (natively supported by the dual port BRAMs) from the integer part of the address and finally the interpolation of the two samples depending on the address fraction:
\begin{equation}
    \text{addr}_i = \text{offset} + i \cdot \text{step} \\
    y[i] = L[k] + (L[k+1] - L[k]) \cdot f
    \label{eq:interpolation_equations}
\end{equation} 
where $k = \lfloor \text{addr}_i \rfloor$ is the integer LUT index (floor of address), $f = \text{addr}_i - k$ is the address fractional part ($0 \le f < 1$), $L[\cdot]$ is the envelope LUT and $y[i]$ represents the interpolated output sample. 
Note that traditional LUT-based systems are a special case of Eq.~\ref{eq:interpolation_equations} where the address increment is fixed to 1.0 and Taylor correction is not needed because the fractional part of the address is always equal to 0. In FIREQ, the linear interpolation function has been optimized to use a single DSP slice, leveraging the pre-adder present in UltraScale+ DSPs to compute the sample difference.
The interpolation stage requires an additional five clock cycles, compared with three cycles for a direct LUT lookup. However, because the DDS path executes in parallel and requires six cycles, the net increase in end-to-end latency is only two cycles (3.2 ns).
As a result of this optimization, memory occupation for an experiment and thus the required LUT memory for each signal generator can be reduced. For this reason, the available memory for each generator in FIREQ is 16k I/Q samples (compared to 32k for QICK full-throughput generators). Because multiple samples must be produced per clock cycle to sustain the DAC throughput (see appendix \ref{app:parallel_env_generation}), interpolation-mode envelopes are replicated across the LUT banks reducing the effective memory available to 1k I/Q samples.
To mitigate this effect, FIREQ generators allow a mix of envelopes to be stored either as-is or with the intention to be used for interpolation, to minimize memory occupation of an experiment, and the interpolation feature supports envelopes with even and odd symmetry, requiring only half of the reference envelope samples to be stored in memory.

\subsubsection{Resource-utilization comparison}
Table~\ref{tab:area_comparison} reports the total resource usage (Logic LUTs, LUTRAMs, SRLs, FFs, BRAM, URAM and DSP)  of the different systems. HiSEP-Q 2.0 is excluded because we were unable to obtain a valid ZCU216 implementation from the publicly available repository. All systems were built from their openly available repositories, using either the recommended build or, when no explicit recommendation was provided, the configuration inferred to be the standard build. At the time of writing, these are: "MultiCoreSoc" for RISC-Q, "qubic" for QubiC and "qick\_tprocv2\_216\_standard" for QICK.

A direct comparison of total device utilization is insufficient because the frameworks differ in supported qubit count, control architecture, and inclusion of auxiliary peripherals (like DDR controllers in QICK).
We therefore devised normalized metrics to estimate the resources needed to control a single qubit, based on the sum of the resources used for the generation and acquisition of drive and readout pulses. When a hardware block can be shared across multiple qubits (like the readout generators and acquisition IPs in QICK), only the corresponding fractional resource cost is included. 
Finally, resources are normalized using the throughput in GHz of the relative IP/module, which is halved if it works with AM instead of I/Q modulation. 
In particular, FIREQ generation and acquisition resources are calculated using one generation IP, one acquisition IP, half the resources for the crossbar and the frequency multiplexing IP. For QICK, one full-speed generator (axis\_signal\_gen\_v6\_0), one eighth of a multiplexed quarter-speed generator (axis\_sg\_mixmux8\_v1) and one eighth of a multiplexed acquisition IP (axis\_pfb\_readout\_v4\_0) were used for the estimation. RISC-Q has AM modulated generators (PulseGenerator + DualClockRam), of which two were counted, and a single acquisition subsystem (ReadoutDecoder + DemodCarrierGenerator). Because the QubiC design hierarchy does not allow an unambiguous decomposition of per-qubit resources, it is excluded from the normalized comparison. 

The resulting normalized metrics are reported in Table~\ref{tab:per_qubit_area}. \mbox{FIREQ} maintains a low Block RAM (BRAM) footprint comparable to RISC-Q, which relies on a CORDIC-based carrier generation approach rather than memory-intensive Direct Digital Synthesizers (DDS). The inclusion of the envelope interpolation feature in the signal generator leads, as expected, to a slight increase in DSP block utilization relative to other architectures, however, DSP slices represent a less constrained resource on the FPGA. 
Similarly, LUT and FF consumption remains comparable across all evaluated solutions and is non-critical given the large logical capacity of the target device. Consequently, \mbox{FIREQ} presents a competitive alternative to existing quantum control architectures, while maintaining a high level of flexibility and reconfiguration capability.

\subsection{Software}
The FIREQ software stack combines a high-level user interface with an extensible and performance-oriented execution architecture. 
The framework adopts a client–server architecture with a dedicated driver layer, separating high-level experiment management from low-level hardware execution, as shown in Fig. \ref{fig:FIREQ_firmware_and_client_server}(b).

This client–server organization is also adopted by other RFSoC-based quantum-control platforms, where it separates host-side experiment management from device-side hardware execution~\cite{stefanazzi2022qick,carobeneQibosoqOpensourceFramework2025}.
Although our framework follows this general architectural paradigm, it introduces several key structural improvements aimed at reducing experiment runtime through a streamed data acquisition system and by offloading most of the computational complexity from the RFSoC to the host computer. 
The software stack is built on PYNQ and includes peripheral drivers and abstraction layers that expose a uniform interface to the underlying hardware. This organization allows new IP blocks and firmware revisions to be integrated without modifying the higher software layers.

The FIREQ client–server software follows a producer–consumer model that enables experiment execution and data transfer to proceed concurrently. Communication between client and server is implemented over TCP.
The client allows experiments to be defined at a high level of abstraction and automatically validates and translates them into low-level hardware configurations. 
The client translates the experiment description into a YAML-compatible hierarchical configuration structure that mirrors the organization of the FIREQ peripherals. Control messages are serialized using MessagePack and transmitted to the server over TCP.
Consequently, the server performs only lightweight validation, firmware configuration, experiment execution, and acquisition handling, while more computationally intensive processing remains on the host, minimizing computational overhead on the RFSoC.

Sweep experiments, which are common in qubit calibration and characterization, can require tens to hundreds of millions of iterations when multiple parameters and repeated shots are combined. Their execution time is therefore highly sensitive to software and configuration overhead. To reduce this overhead, FIREQ implements a dependency-aware update mechanism that recomputes and applies only the configuration changes required by the parameters modified between consecutive iterations.

\subsubsection{Client}
The client software provides a high-level abstraction through a dedicated API that allows users to define  quantum experiments. Experiments are described through a declarative, compiler-like syntax rather than by directly constructing low-level hardware configuration structures. 
Experiment parameters can be expressed either in physical units or in hardware-specific quantities such as clock cycles, according to user needs. 
The API converts the experiment description into a hierarchical configuration dictionary, which is then subjected to consistency and feasibility checks before transmission to the server. These checks verify both the physical feasibility of the requested pulses with respect to hardware constraints and the logical consistency of the experiment timeline. 
Because most validation is performed on the client, the server, on the RFSoC, can focus on hardware configuration, real-time execution, and acquisition.

\subsubsection{Server}
The server is organized into three threads with separate responsibilities:
\begin{enumerate}
    \item Sender thread: sends packets to the client. Packets consist of a MessagePack-serialized header followed, when required, by an optional  binary data payload. 
    \item Receiver thread: receives control packets from the client and places them in a shared input queue. 
    \item Execution thread: consumes incoming commands, configures the hardware, executes experiments, and handles runtime errors. 
\end{enumerate}
The three-thread organization implements a producer–consumer model. The receiver thread enqueues incoming commands, the execution thread is the only component that accesses the hardware, and the sender thread transmits results and status messages. Restricting hardware access to the execution thread avoids race conditions. This organization allows command reception, experiment execution, and result transmission to proceed concurrently.
The acquisition pipeline is designed to overlap experiment execution with data transmission.
The execution thread groups multiple shots into acquisition batches whose size is determined by the generated data volume and the available acquisition memory. This reduces per-shot software overhead. Acquisition results are placed in the outgoing queue using zero-copy memory views and transmitted as binary chunks. A compact header identifies the originating acquisition IP and the data format. Reassembly of the binary chunks into complete floating-point arrays is offloaded to the client. Small metadata headers and progress-update messages provide the information required to reconstruct the acquired dataset and track sweep execution.

\subsubsection{Firmware abstraction layers}
FIREQ was designed to support the integration of new hardware IPs without requiring substantial changes to the upper software layers.
To this end, the server uses a tree-based, driver-centric abstraction model. Each peripheral is represented by a software object exposing a standardized configuration interface, while a dependency-resolution mechanism propagates configuration changes across related objects. The abstraction stack is implemented through two Python packages  FIREQ\_LL\_API and FIREQ System. 
The FIREQ\_LL\_API provides a low-level PYNQ-based interface. It includes peripheral drivers and handles system initialization, including overlay loading, RF-clock configuration, parsing of the hardware handoff file, and reconstruction of the hardware connectivity graph from which global parameters such as clock topology and clock frequencies are extracted.
\begin{figure}
    \centering
    \includegraphics[width=1.0\linewidth]{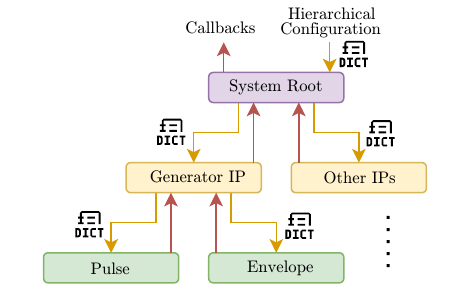}
    \caption{Example of the FIREQ system tree, showing how the hierarchical configuration is applied and the resulting return callbacks.}
    \label{fig:config_example}
\end{figure}
The FIREQ System package builds upon the low-level API and provides wrapper classes that expose the functionality of each peripheral. Each IP instance is represented as a node in a hierarchical system tree. Configuration updates are expressed through hierarchical configuration dictionaries that are recursively applied from the system root to the affected peripheral nodes, or, more generally, system nodes, as shown in Fig.~\ref{fig:config_example}. In fact, nodes not only represent peripherals, but can also represent peripheral objects that require behavior that cannot be represented by simple configuration parameters. These supporting node types can be dynamically created and modified by the configuration itself if a certain node supports the instantiation of a child node.
This configuration can not only change HW parameters (such as the drive or readout modulation frequency of a specific generator IP instance) but can also create children within certain peripheral that model certain properties, allowing for even more flexible reconfiguration options. 
For example, in FIREQ signal generator peripheral nodes, pulses and envelopes are represented as children nodes and, once created, their parameters can be modified like the parameters of any other peripheral.
The same configuration mechanism supports sweep expressions associated with node parameters. During configuration, these expressions are compiled into a list of callback functions annotated with their corresponding update expression and execution cost, which is a pre-calculated metric attached to the callback function in the node class.
During a sweep, only callbacks affected by the variables modified at the current iteration are executed. This avoids reconstructing and reapplying the complete hardware configuration at every sweep point. 
Moreover, an optimal variable execution order is determined before running the sweep using the callback execution costs, to minimize the software overhead.
Dependencies between configuration objects are explicitly registered with the system root. When a parameter changes, the dependency resolver identifies the affected objects and invokes only the corresponding update functions before the next experiment iteration. This mechanism limits configuration recomputation to the portions of the hierarchy that are actually affected by the sweep variable.

\section{SYSTEM VALIDATION AND EXPERIMENTAL RESULTS} \label{results}

\subsection{Instrument characterization}
The instrument was characterized in terms of phase noise, noise spectral density and inter-channel timing skew.
\begin{table}[ht]
    \centering
    \small 
    \caption{Phase noise measurements in dBc/Hz for two tones (513.63 and 4300 MHz), using the XM655 balun add-on card to extract the RF output.}
    \label{tab:phase_noise_meas}
    \begin{NiceTabular}{|c|c|c|c|}
        \hline
        \textbf{Gain} & \textbf{Offset [Hz]} & \textbf{513.63 MHz} & \textbf{4300 MHz} \\
        \hline
        \multirow{5}{*}{\textbf{0.005}} 
        & 100  & -88.67 & -76.47 \\ \cline{2-4}
        & 1k   & -94.35 & -82.86 \\ \cline{2-4}
        & 10k  & -94.85 & -84.23 \\ \cline{2-4}
        & 100k & -94.54 & -85.18 \\ \cline{2-4}
        & 1M   & -94.72 & -85.05 \\
        \hline
        \multirow{5}{*}{\textbf{0.05}} 
        & 100  & -89.07  & -77.81 \\ \cline{2-4}
        & 1k   & -99.10  & -86.12 \\ \cline{2-4}
        & 10k  & -103.10 & -89.47 \\ \cline{2-4}
        & 100k & -112.55 & -102.68 \\ \cline{2-4}
        & 1M   & -114.75 & -104.77 \\
        \hline
        \multirow{5}{*}{\textbf{0.5}} 
        & 100  & -89.65  & -77.29 \\ \cline{2-4}
        & 1k   & -99.43  & -85.39 \\ \cline{2-4}
        & 10k  & -103.32 & -89.17 \\ \cline{2-4}
        & 100k & -116.68 & -107.16 \\ \cline{2-4}
        & 1M   & -132.16 & -119.20 \\
        \hline
    \end{NiceTabular}
\end{table}
\begin{figure}[ht]
    \centering
    \includegraphics[width=0.99\linewidth]{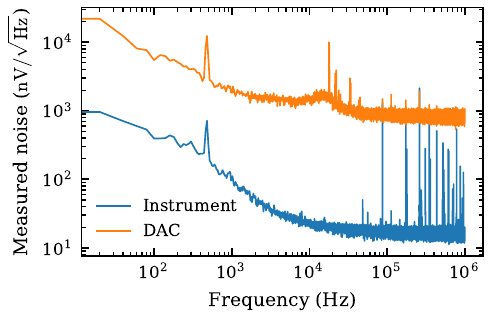}
    \caption{Noise spectral density of the 50-ohm terminated instrument and of a single branch of the DAC's differential output between DC and 1 MHz, analyzed from 8M samples taken at 8MSps and using the Welch method (Welch window of 400k samples and 50\% overlap).}
    \label{fig:noise_floor}
\end{figure}
Table~\ref{tab:phase_noise_meas} reports the phase noise measurements for different output amplitudes and carrier frequencies. The phase noise was measured with an Agilent Technologies MXA N9020A Signal Analyzer while the RFSoC generated a continuous tone. The RF analog output was accessed through the AMD XM655 add-on card, using the 10 MHz-1 GHz balun path for the low frequency tone and the 4-5 GHz path for the high frequency tone. Measurements were taken at frequency offsets of 100 Hz, 1 kHz, 10 kHz, 100 kHz and 1 MHz, using a resolution bandwidth (RBW) of 1 Hz, 10 Hz, 100 Hz, 1 kHz and 10 kHz, respectively.
At the 100 Hz offset, the measured phase noise shows only weak dependence on the programmed gain. At larger offsets, the measurements at low output amplitudes approach the instrument noise floor, resulting in nearly constant reported values.
For the highest output amplitude, the measured phase noise at a 1 MHz offset is below -116 dBc/Hz, the level reported by Van Dijk et al.\cite{ddsdesignqubit} as compatible with high-fidelity gate operation.
%
\begin{table}[ht]
    \centering
    \small
    \caption{Noise spectral density of the 50-ohm terminated instrument and of a single branch of the DAC's differential output at specific frequencies.}
    \label{tab:noise_markers}
    \begin{NiceTabular}{|c|c|c|}
        \hline
        \textbf{Offset [Hz]} & \textbf{Instrument [nV/$\sqrt{\text{Hz}}$]} & \textbf{DAC [nV/$\sqrt{\text{Hz}}$]} \\
        \hline
        1k   & 99.61 & 1909.53 \\
        \hline
        10k  & 23.05 & 1530.58 \\
        \hline
        100k & 15.25 & 884.41  \\
        \hline
        1M   & 16.94 & 948.85  \\
        \hline
    \end{NiceTabular}
\end{table}
\begin{figure*}[p]
	\centering
	\includegraphics[width=1.0\linewidth]{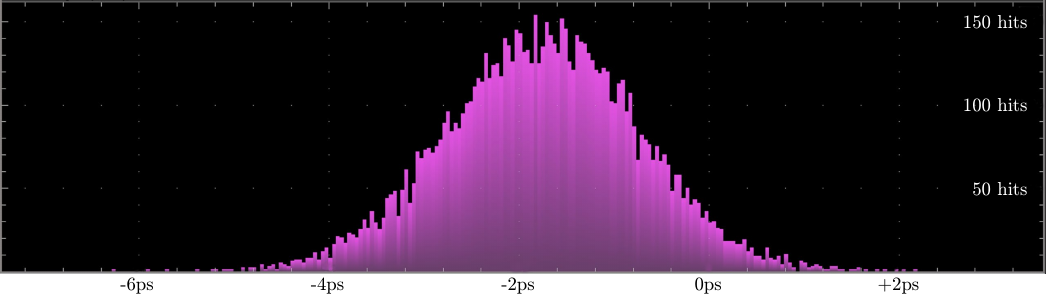} 
\caption{Measurement of channel to channel skew. Performed through the generation of two simultaneous pulses on two different channels and measuring their skew at the 50\% rise mark for a train of 9000 pulses.}
\label{fig:skew_histogram}
\end{figure*}
The low frequency Noise Spectral Density (NSD) was estimated from two 1s acquisitions performed with an Agilent MSO6104A 1 GHz, configured with a 50-ohm input. The oscilloscope was first connected to a 50-ohm termination to determine its intrinsic noise floor and was then connected to one branch of the DAC differential outputs. The acquired data were analyzed using Welch's method. The resulting spectra are shown in Fig.~\ref{fig:noise_floor}, while representative values are reported in Table~\ref{tab:noise_markers}.
As expected, the DAC-output spectrum lies above the oscilloscope noise floor and exhibits an approximately 1/f profile characteristic of pink noise over part of the measured band. 
A direct comparison with commercial instruments was not available. However, the measured noise density at 100 kHz and 1 MHz (in the order of 1000 $\text{nV}/\sqrt{\text{Hz}}$) is higher than that reported for other open-source RFSoC-based solutions using the same RF-DAC together with dedicated low-noise bias circuitry connected through a bias tee\cite{manarat}. Indeed, in Maranat\cite{manarat}, the authors achieved  100.5 $\text{nV}/\sqrt{\text{Hz}}$ at 100 kHz and 13.8 $\text{nV}/\sqrt{\text{Hz}}$ at 1 MHz.
Because this measurement primarily characterizes the analog DAC output path rather than the FPGA processing chain, the excess noise is unlikely to originate from the digital firmware. A possible explanation is a common-mode contribution on the differential DAC output. If this interpretation is confirmed, direct driving of a single-ended DC-flux line would require a low-noise differential-to-single-ended stage rather than connection to a single branch of the RFSoC differential output.
%
Finally, we characterized the relative timing skew between the two drive channels. The corresponding DACs belong to different RFSoC tiles and are synchronized automatically using Multi Tile Synchronization (MTS). The DAC outputs were directly connected to a four-channel DPO714AX oscilloscope, with an analog bandwidth of 25 GHz and a sampling rate of 125 GSps. 
A sequence of 9000 simultaneous pulse pairs was generated. The relative timing offset between CH1-CH2 and CH3-CH4 was measured at the 50\% transition point and summarized in the histogram of Fig.~\ref{fig:skew_histogram}.
The resulting distribution shows an absolute mean skew of approximately 1.8 ps 
and a standard deviation of 0.84 ps. 
These values are substantially lower than the 107 ps timing resolution of the RF DAC.
\subsection{Qubit measurements}
\begin{figure*}[p]
	\centering
	\includegraphics[width=1.0\linewidth]{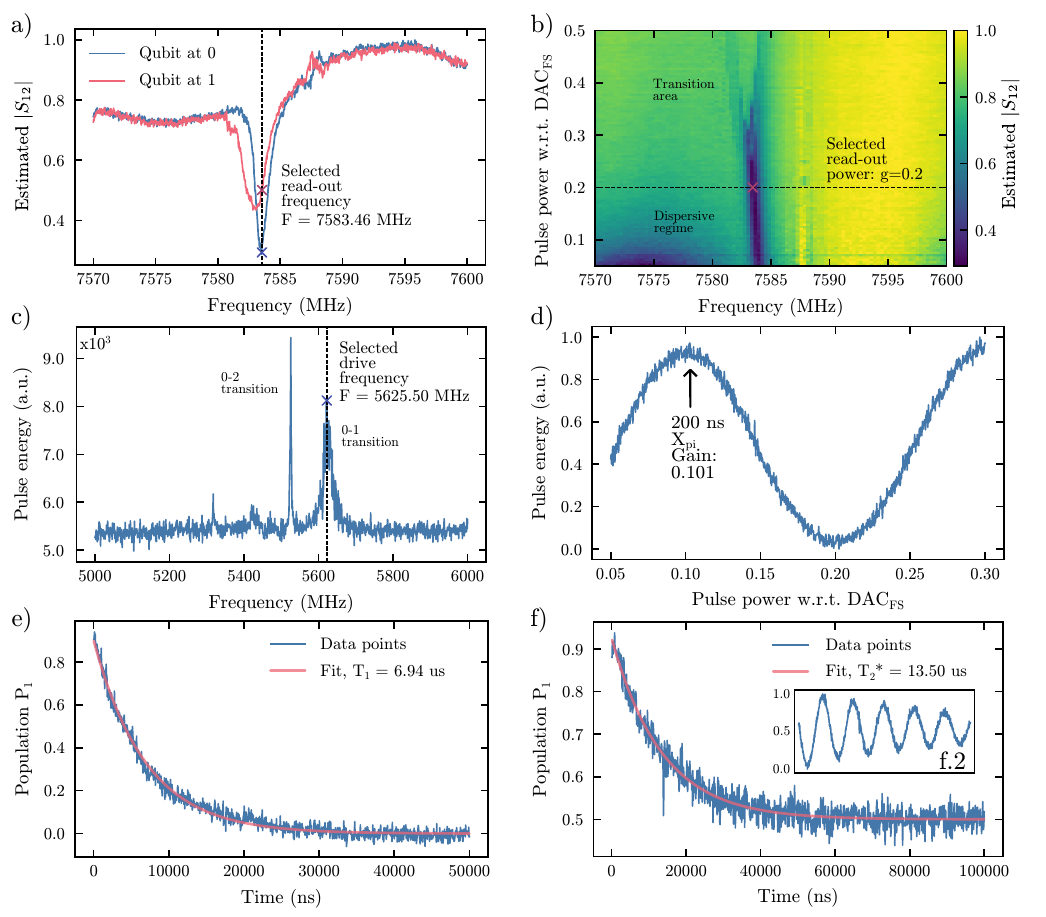} 
\caption{Qubit characterization experiments performed at the University of Milano Bicocca. (a) Readout spectroscopy with the qubit prepared in state $\ket{0}$ and $\ket{1}$ and the selected readout frequency, (b) readout punch-out and the selected readout power, (c) qubit spectroscopy showing the $\ket{0}\rightarrow\ket{1}$ and $\ket{0}\rightarrow\ket{2}$ transitions and the initial drive frequency selected, (d) Rabi oscillations of a 200 ns drive pulse and the $X_{\pi}$ pulse calibration, (e) $T_1$ experiment, (f) Ramsey experiment to extract the $T_2^*$, (f.2) detuned Ramsey measurement used for drive-frequency refinement.}
\label{fig:qubit_measurements}
\end{figure*}
\begin{table*}[ht]
    \centering
    \caption{Description of experiments.}
    \label{tab:exp_specs}
    \begin{NiceTabular}{|p{2.3cm}|p{4cm}|p{3.2cm}|p{3.2cm}|}
        \hline
        \textbf{Experiment}   & \textbf{Description} & \textbf{Sweep}     & \textbf{Iterations} \\
        \hline
        Resonator spectroscopy at $\ket{0}$ (Fig~\ref{fig:qubit_measurements}.a) & 
        1us square readout pulse & 
        Readout frequency, 1000 averaged shots per point & 
        1000 points between 7570 and 7600 MHz\\
        \hline
        Resonator punch-out (Fig~\ref{fig:qubit_measurements}.b) &
        1us square readout pulse &
        Readout frequency and gain (pulse amplitude), 1000 shots per point &
        10000 total points (100 x 100) between 0.05 and 0.5 for the gain and 7570 and 7600 MHz for the frequency\\
        \hline
        Drive (qubit) spectroscopy (Fig~\ref{fig:qubit_measurements}.c) &
        6us drive pulse followed by a 1us readout pulse &
        Drive pulse frequency, 1000 shots per point &
        1000 points between 5000 and 6000 MHz \\
        \hline
        Rabi oscillations (Fig~\ref{fig:qubit_measurements}.d)&
        200ns drive pulse followed by a 1us readout pulse &
        Drive pulse amplitude (gain), 1000 shots per point &
        1000 points between 0.05 and 0.5 \\
        \hline
        Resonator spectroscopy at $\ket{1}$ (Fig~\ref{fig:qubit_measurements}.a)&
        $X_{\pi}$ drive followed by a 1us readout pulse &
        Readout frequency, 1000 averaged shots per point & 
        1000 points between 7570 and 7600 MHz\\
        \hline
        $T_1$ (Fig~\ref{fig:qubit_measurements}.e)&
        $X_{\pi}$ drive, variable delay followed by a 1us readout pulse&
        Delay between the drive and readout pulses, 1000 shots per point&
        1000 delays between 10ns and 50us \\
        \hline
        Ramsey experiment (Fig~\ref{fig:qubit_measurements}.f)&
        Two $X_{\pi/2}$ drive pulses with an increasing delay between them, followed by a 1us readout pulse &
        Delay between the two pulses, 1000 shots per point &
        1000 inter-gate delays between 100ns and 100us \\
        \hline
    \end{NiceTabular}
\end{table*}
\begin{table}[ht]
    \centering
    \caption{Outcome of the calibration experiments performed at the University of Milano-Bicocca.}
    \label{tab:calibration_outcome}
    \begin{NiceTabular}{|p{2cm}|p{3cm}|p{3.3cm}|}
        \hline
        \textbf{Parameter} & \textbf{Calibrated/ measured value} & \textbf{Comments} \\
        \hline
        Readout pulse & F = 7583.46 MHz\newline G = 0.20\newline L = $1 \mu s$ & F chosen from the resonating frequency, no further optimization was performed\\
        \hline
        $X_{\pi}$ pulse & F = 5625.213 MHz\newline  G = 0.101\newline  L = 200 ns & \\
        \hline
        $X_{\pi/2}$ pulse & F = 5625.213 MHz\newline  G = 0.0505\newline  L = 200 ns & Gain was halved w.r.t. the $X_{\pi}$ pulse\\
        \hline   
        $T_2^*$ & $13.50 \mu s$ & \\
        \hline
        $T_1$ & $6.94 \mu s$ & \\
        \hline
    \end{NiceTabular}   
\end{table}
FIREQ was experimentally validated on a superconducting qubit at the cryogenic facilities of the University of Milano Bicocca. The validation included the calibration of the drive and readout parameters and measurement of the characteristic relaxation and dephasing times $T_1$ and $T_2^*$.
The readout parameters were calibrated through a readout resonator spectroscopy and punch-out. These experiments determine the resonator response near the $\ket0$ state and identify an operating readout amplitude compatible with the dispersive regime. 
Fig.~\ref{fig:qubit_measurements}(a) shows the resonator spectroscopy used to select the readout frequency. Fig.~\ref{fig:qubit_measurements}(b) reports the punch-out measurement used to characterize the power-dependent resonator response and select the readout gain that provides adequate SNR while remaining in the dispersive regime.
Then, an initial value for the drive frequency was selected through a qubit spectroscopy (Fig~\ref{fig:qubit_measurements}c), which resolves the drive frequency of the $\ket{0}\rightarrow\ket{1}$ transition.
The $X_{\pi}$ pulse amplitude was calibrated from Rabi oscillations using a fixed pulse duration of 200 ns, Fig.~\ref{fig:qubit_measurements} (d). The measured readout response was recorded as a function of the drive-pulse amplitude. The first maximum corresponds to a $\pi$-rotation that transfers the qubit from $\ket{0}$ to $\ket{1}$. Halving this gain also provides the $X_{\pi/2}$ pulse.
The drive frequency was then fine-tuned using the Ramsey experiment, as shown in Fig.~\ref{fig:qubit_measurements} (f.2), whose oscillation frequency is given by $\Delta F = |F_{\text{qubit}} - F_{\text{drive}}|$.
Finally, the $T_1$ and $T_2^*$ times were extracted by fitting the corresponding exponential curves, yielding $T_1 = 6.94 \mu s$ and $T_2^* = 13.50 \mu s$ as shown in Fig.~\ref{fig:qubit_measurements} (e) and (f).
Table~\ref{tab:exp_specs} summarizes the experimental sequences and sweep parameters, while Table~\ref{tab:calibration_outcome} reports the calibrated drive and readout settings.
The calibration procedures in Table~\ref{tab:exp_specs} combine parameter sweeps with repeated-shot acquisition, resulting in total shot counts ranging from millions to tens of millions. 
The execution path in FIREQ was therefore designed to minimize firmware, software, and communication overhead. 
\begin{table}[ht]
  \centering
  \caption{Assessment of the software overhead for the qubit calibration experiments.}
  \label{tab:execution_time_comparison}
  
  \begin{NiceTabular}{|l|c|c|c|}
    \hline
    \Block{2-1}{\textbf{Experiment}} & \Block{1-2}{\textbf{Time [s]}} & & \Block{2-1}{\shortstack{\textbf{Overhead}\\\textbf{[\%]}}} \\
    \cline{2-3}
    & \textbf{FIREQ} & \textbf{HW Time} & \\
    \hline
    Resonator Spectroscopy & 3.7  & 3.0 & 23\%\\ \hline
    Resonator Punch-out    & 37.8  & 30.0 & 26\%\\ \hline
    Qubit Spectroscopy    & 10.8  & 10.0 & 8\%\\ \hline
    Rabi experiment        & 70.8  & 70 & 1\%\\ \hline
    Ramsey experiment      & 101.5  & 100 & 1.5\%\\ \hline
    T1 experiment          & 60.8 & 60 & 1.3\% \\
    \hline
  \end{NiceTabular}
\end{table}
As shown in Table~\ref{tab:execution_time_comparison}, the software overhead varies depending on the experiment. Because the per-iteration software latency is essentially constant, shorter experiments, such as the resonator spectroscopy and punch-out which take only $3\mu s$ per iteration, incur a higher relative penalty (up to 26\% overhead), whereas longer routines effectively amortize this fixed cost down to roughly 1\%. \\
The results of the qubit characterization experiments obtained with FIREQ are comparable with the ones obtained by our colleagues at the University of Milano Bicocca using QICK\cite{stefanazzi2022qick}.

\section{CONCLUSIONS AND FUTURE WORK}
We presented the FPGA Instrumentation for Readout and Qubit control (FIREQ), an RFSoC-based framework for the control and readout of superconducting qubits. 
FIREQ directly synthesizes precisely timed drive and frequency-multiplexed readout RF pulses with programmable envelopes at frequencies up to 9.3 GHz, with a pulse-duration resolution of 107 ps and an event-timing resolution of 1.7 ns. The specialized signal-generation and acquisition architectures were designed to reduce on-chip memory requirements, particularly BRAM utilization. The normalized comparison with representative open-source RFSoC frameworks shows that FIREQ achieves a low BRAM footprint while maintaining full-rate I/Q generation and acquisition.
The RF output was characterized in terms of phase noise, noise spectral density, and inter-channel timing skew, while end-to-end operation was validated through standard calibration and coherence measurements on a superconducting qubit. The resulting measurements yielded $T_1 = 6.94 \mu s$ and $T_2^* = 13.50 \mu s$. The client–server execution stack was also exercised over calibration sweeps comprising millions of repeated shots; the relative software and communication overhead decreases from up to 26\% for the shortest sequences to approximately 1–1.5\% for longer experiments.

Future work will extend FIREQ to larger qubit counts, incorporate fast and slow flux-bias control channels, and support synchronized multi-board operation. Further work will investigate lower-resource implementations of the generation and acquisition peripherals, including architectures suitable for migration toward ASIC-based control electronics while preserving low-latency operation. 


\section*{Author Declarations}
\subsection*{Conflict of Interest}
The authors have no conflicts to disclose.
\section*{Data availability}
The FIREQ software and documentation associated with this work are archived in
versioned Zenodo records. The FIREQ client software release 0.1.0 is available
in~\cite{fireq_client_zenodo}. The FIREQ server-side software release
0.1.0, including the executable RFSoC bitstream required to reproduce the
reported setup, is available in~\cite{fireq_server_zenodo}.
The documentation release 0.1.0 is available 
in~\cite{fireq_docs_zenodo}.\\
The public development repositories and user documentation are accessible from
the FIREQ documentation website or GitHub. The experimental data that support the findings
of this study are available from the corresponding author upon reasonable
request.

\section*{Acknowledgements}
This work was supported in part by AMD under the  \href{https://www.amd.com/en/corporate/university-program.html}{AMD University Program}. This work is also supported by QUART\&T, a project funded by the Italian Institute of Nuclear Physics (INFN) within the Technological and Interdisciplinary Research Commission (CSN5) and Theoretical Physics Commission (CSN4) and by PNRR MUR projects PE0000023-NQSTI and CN00000013-ICSC.


%
%

%


\appendix

\section{Parallel envelope generation with interpolation}
\label{app:parallel_env_generation}

Because the RF DAC operates at approximately 9.3 GSps, the FPGA datapath cannot generate samples serially at the DAC sampling rate. FIREQ signal generators therefore produce N=16 samples in parallel at each generator clock cycle, corresponding to a generator clock frequency of approximately $9.3/16\approx581 MHz$.
Fundamentally, parallelization (unfolding) increases the resources used, leading to the same logic being duplicated N times to provide N samples at the output. In the following, each parallel sample-generation path is referred to as a lane. 
Producing N samples per cycle requires N simultaneous LUT reads. Since a dual-port BRAM provides at most two independent reads per cycle, at least N/2 memory banks are required. For conventional non-interpolated envelopes, these banks do not replicate the waveform data: the envelope samples can instead be statically partitioned across banks. 
For a conventional LUT lookup with unit address increment, lane j always accesses samples j, j+N, j+2N, etc.... The complete envelope, based on Eq.~\ref{eq:interpolation_equations} can therefore be interleaved across the memory banks, with each lane storing only the samples that it can address. Consequently, the aggregate capacity of the banks remains equal to the logical envelope capacity, no waveform-data replication is required.
With interpolation, the address increment is programmable and generally non-integer. As a result, the two LUT samples required to reconstruct each output sample cannot be assigned statically to a single lane-specific bank. Each of the N parallel interpolators requires simultaneous access to two adjacent reference samples, i.e., 2N read accesses per generator cycle. With dual-port BRAMs, this requires N replicated memory banks. Because each bank must contain the same reference-envelope data, the physical memory is replicated N times rather than partitioned across the lanes. The effective logical envelope capacity is therefore reduced by a factor N. For N=16, the 16 kSample capacity available to conventional envelopes becomes 1 kSample for interpolated reference envelopes.

\section*{References}
\bibliography{references}

\end{document}